# Real Time Relativity

Craig M. Savage, Antony C. Searle, Lachlan McCalman,
*Department of Physics, Faculty of Science, The Australian National University,
ACT 0200*
www.anu.edu.au/Physics/Savage/RTR

## A virtual world

Real Time Relativity is a computer program that allows the user to fly through a virtual world governed by relativistic physics. The experience is that of flying a relativistic rocket in a 3D computer game. It is available for Windows systems only and may be downloaded from the Real Time Relativity web site [1].

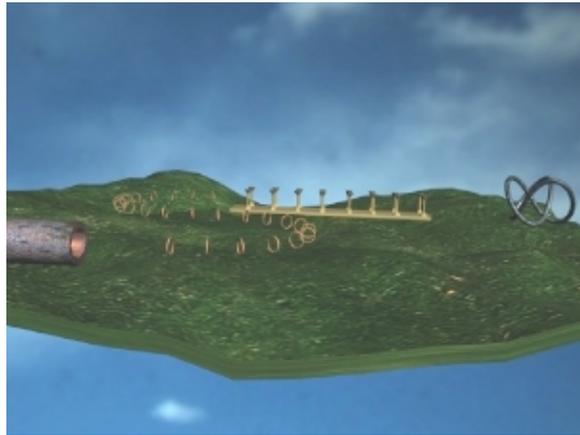

The user controls a "rocket" and sees on their screen the view of a "camera" on the rocket. The rocket may be accelerated and steered, and the camera may be pointed in any direction. Relativistic physics is used to determine how objects in the world frame appear in the flying camera frame.

Real Time Relativity (RTR) uses the fact that video cards provide inexpensive data-parallel processing and are designed to perform rapid arithmetic on four-dimensional vectors. These are usually interpreted as pixel colour values, (red, green, blue, opacity), but in RTR are also photon energy-momentum 4-vectors (see below).

We are interested in the potential uses of RTR in physics education. Our first question is: "Can aspects of relativity be learnt by exploring such a virtual world?" Because relativistic physics is not part of our direct experience, traditional ways of learning it are often abstract and mathematical. But today many people are comfortable in the virtual worlds of computer games, and are used to discovering their "physics" by experimentation. Might students begin to discover relativistic physics by exploring a relativistic virtual world?

## Physics

Real Time Relativity implements relativistic physics such as: aberration, the Doppler shift, and the headlight effect. The 2D screen image is created using the computer graphics technique known as environment mapping, which renders the 3D virtual world onto a 2D cube map. A cube map may be visualised as the 360-degree camera view-field mapped onto the interior surface of a cube enclosing the camera. In fact, the cube map is a data structure in which the image pixels are addressed by line of sight direction [2], rather than by spatial position.

Relativistic aberration is the dependence of the direction of incoming light rays on the relative motion of the camera and the objects from which they originate. Each camera image pixel is formed by photons incident from a particular direction; that is by light with a specific propagation vector in the camera frame. The relativistic physics problem is to find the corresponding vector in the virtual world frame. This vector then addresses the pixel on the cube map that is mapped to the camera pixel. The resulting camera image is displayed on the screen.

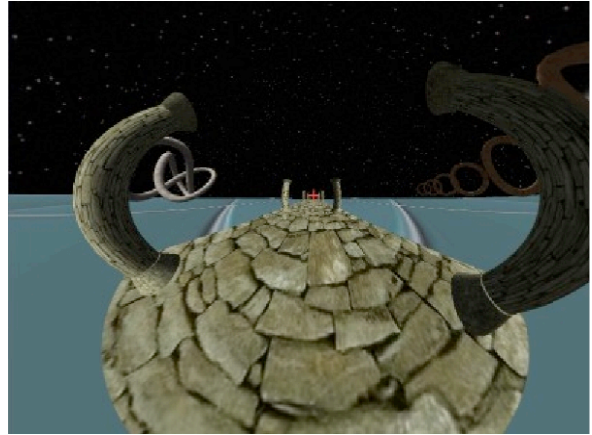

In relativity a photon is represented by its relativistic energy-momentum 4-vector $\mathbf{P} = hf(1,\hat{\mathbf{n}})/c$, where $h$ is Planck's constant, $f$ is the photon frequency, $\hat{\mathbf{n}}$ the unit 3-vector in its propagation direction, and $c$ is the speed of light. The propagation direction in the world frame is found by the Lorentz transformation $L$ of this 4-vector in the camera frame, $\mathbf{P}_C$, into the world frame: $\mathbf{P}_W = L\,\mathbf{P}_C$. The spatial 3-vector component of $\mathbf{P}_W$ is along the photon's propagation direction in the world frame. A 4x4 matrix represents the Lorentz transformation $L$. It is calculated before each frame is rendered, using the current camera velocity, and is then applied to each camera pixel's photon energy-momentum 4-vector. The spatial parts of these vectors address the cube map pixel that is to be rendered to the user's screen. Since they are specifically designed to process 4-vectors in parallel, video card Graphics Processing Units (GPUs) [3] can perform the 4D Lorentz transformation in real time. However, this approach is currently limited to static worlds in which the objects do not move.

The Doppler shift of the photon frequency is given by the ratio of the time components of the photon energy-momentum 4-vector, see above. However, to determine the effect of the Doppler shift on a general colour requires the intensity spectrum for each pixel. But in current implementations the spectrum is specified at just three frequencies; red, green, and blue. Hence a simple interpolation is used to generate the intensity spectrum. This simple approach is a significant limitation of the current version of RTR.

At relativistic velocities aberration concentrates the incident light into a narrow cone centred on the direction of motion. In addition, time dilation increases the photon flux in the camera frame. Overall there is a brightening in the direction of motion, and a darkening away from the direction of motion. The detected intensity scales as the fourth power of the Doppler shift [4]. Again, there are significant limitations on how the resulting large intensity range is rendered to the screen by the current version of RTR.

# Technology

The video card does the graphics work and the Lorentz transformations. The main loop has four major steps. 1) The camera's position, velocity, and orientation are calculated from the user input. 2) Using this information the video card renders the 3D virtual world to a world frame 2D cube map. 3) The GPU Lorentz transforms the scene into the camera frame. 4) The output is displayed on the screen.

RTR displays an 800 by 600 pixel window of 480,000 pixels. Each associated photon 4-vector is Lorentz transformed to find the corresponding world frame cube map pixel, which is then Doppler and intensity shifted. A typical PC setup can display 50 frames per second, corresponding to 24 million Lorentz transformations per second. This is well within the capabilities of even low end GPUs, and hence the conventional graphics process of the cube map rendering limits the overall performance, not the relativistic calculations.

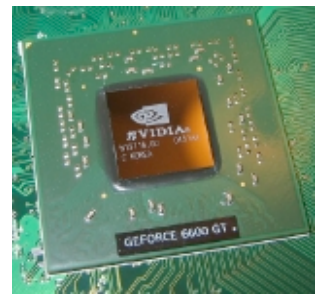

GPUs have been following a super Moore's law, doubling in processing power every 6 months, compared to every 18 to 24 months for CPUs [5]. This is driven by the demand for parallel computing from the gaming community. For example, the Xbox 360 GPU has 48 processors running at 500MHz, each capable of a floating point 4-vector operation per cycle, giving nearly 100 GFlops, compared to a few GFlops for a Pentium 4 [6]. The main catch is that GPUs do data-parallel computing, in which the same operation is repeated on a data array. Nevertheless, computational scientists are developing algorithms that harness the commodity processing power of GPUs for useful tasks, such as solving PDEs; a field called "General Purpose Computing on GPUs" [5,7].

The type of GPU program RTR uses is called a pixel-shader. This is a small program that performs per-pixel operations, after the vertices in the scene have been manipulated, and before the pixel is output to the screen. This is an ideal point in the render pipeline to implement the relativistic optics transformations for two reasons. Firstly, as the operations are per-pixel, the geometry of the scene is irrelevant: more complicated geometry has no effect on the cost of the pixel shader. Secondly, the GPU has built in support for 4-vector arithmetic, making relativistic calculations easy to code and fast to run.

RTR is programmed using Microsoft's DirectX 9 API, so that it is independent of the particular GPU available. DirectX 9 includes the C++ like High Level Shader Language [8], in which the pixel shader is written. Consequently, it is only available on Windows computer systems.

# Past and future

RTR builds on previous relativistic computer graphics work at ANU, including the Through Einstein's Eyes project [9], which used the Backlight program [10,11]. The only other interactive relativistic graphics systems that we are aware of were developed at the University of Tübingen. An early one used parallel CPUs [12], and the later one a GPU [13]. In the latter the user rides a bicycle through the streets of a virtual city. It is on exhibit in German museums [14].

RTR works because video cards provide data-parallel processing of 4-vectors. As algorithms, GPUs and CPUs increase in power students will eventually be able to interact with a dynamic virtual relativistic world. Might this be harnessed to improve students' understanding of relativity?